\documentclass[lettersize,journal]{IEEEtran}
\usepackage{amsmath,amsfonts}
\usepackage{algorithmic}
\usepackage{algorithm}
\usepackage{array}
\usepackage[caption=false,font=normalsize,labelfont=sf,textfont=sf]{subfig}
\usepackage{textcomp}
\usepackage{stfloats}
\usepackage{url}
\usepackage{verbatim}
\usepackage{graphicx}
\usepackage{cite}
\usepackage[pagewise]{lineno}
\usepackage{makecell}
\usepackage{xcolor}
\usepackage{soul}
\newcommand{\revision}[1]{\textcolor{black}{#1}}
\begin{document}

\title{Neural Architecture Search generated Phase Retrieval Net for Real-time Off-axis Quantitative Phase Imaging}

\author{Xin~Shu, Mengxuan~Niu, Yi~Zhang, Wei~Luo, and Renjie~Zhou,
\thanks{This work was supported in part by Hong Kong Innovation and Technology Fund under Grant ITS/148/20, Grant ITS/178/20FP, and Grant ITS/229/23FP; in part by the Croucher Foundation under Grant CM/CT/CF/CIA/0688/19ay; in part by the Research Grant Council of Hong Kong SAR under Grant 14209521; and in part by the National Natural Science Foundation of China/Research Grants Council of Hong Kong Joint Research Scheme under Grant N\_CUHK431/23. \textit{(Corresponding author: Renjie Zhou.)}}%
\thanks{Xin Shu and Renjie Zhou are with the Department of Biomedical Engineering, Chinese University of Hong Kong, Shatin, New Territories, Hong Kong SAR, China (email: shuxin@link.cuhk.edu.hk, rjzhou@cuhk.edu.hk).}%
\thanks{Mengxuan Niu is with the School of Mechanical and Electrical Engineering, Shenzhen Polytechnic University, Shenzhen, Guangdong, China (email: mengxuanniu@szpu.edu.cn).}
\thanks{Yi Zhang is with the Institute of Data Science, University of Hong Kong, Hong Kong Island, Hong Kong SAR, China (email: yizhang101@connect.hku.hk).}
\thanks{Wei Luo is with the College of Biomedical Engineering and Instrument Science, Zhejiang University, Hangzhou, China (email: luo.wei@zju.edu.cn).}
}

\maketitle

\begin{abstract}
In off-axis Quantitative Phase Imaging (QPI), artificial neural networks have been recently applied for phase retrieval with aberration compensation and phase unwrapping. However, the involved neural network architectures are largely unoptimized and inefficient with low inference speed, which hinders the realization of real-time imaging. Here, we propose a Neural Architecture Search (NAS) generated Phase Retrieval Net (NAS-PRNet) for accurate and fast phase retrieval. NAS-PRNet is an encoder-decoder style neural network, automatically found from a large neural network architecture search space through NAS. By modifying the differentiable NAS scheme from SparseMask, we learn the optimized skip connections through gradient descent. Specifically, we implement MobileNet-v2 as the encoder and define a synthesized loss that incorporates phase reconstruction loss and network sparsity loss. NAS-PRNet has achieved high-fidelity phase retrieval by achieving a peak Signal-to-Noise Ratio (PSNR) of $36.7$ dB and a Structural SIMilarity (SSIM) of $86.6\%$ as tested on interferograms of biological cells. \revision{Notably, NAS-PRNet achieves phase retrieval in only $31$ ms, representing }$15\times$ \revision{speedup over the most recent Mamba-UNet with only a slightly lower phase retrieval accuracy.} 
\end{abstract}

\begin{IEEEkeywords}
Phase retrieval, neural architecture search, quantitative phase imaging, real-time reconstruction.
\end{IEEEkeywords}
\vspace{-0.3cm}
\section{Introduction}
Quantitative Phase Imaging (QPI) has been widely applied to biomedical imaging and material metrology. Among all approaches off-axis interferometry-based QPI methods (off-axis QPI) can offer high speed phase imaging due to their single-shot image acquisition feature~\cite{QPIinBME, niu2024compact, ling2020high}. However, retrieving the phase map from a recorded fringe pattern or interferogram requires several image processing steps, which means the specimens are usually not observed in real-time. In conventional approaches, the steps include (i) retrieving the wrapped phase (valued between $-\pi$ to $\pi$) from the object interferogram (e.g., the Fourier transform-based method~\cite{takeda1982fourier}); (ii) calibrating the phase or compensating the phase aberration by using an additional interferogram captured in a sample-free region~\cite{ferraro2003compensation}; and (iii) unwrapping the phase (e.g., the Goldstein algorithm~\cite{goldstein1988satellite}). Among these steps, phase unwrapping is the most time-consuming. To expedite phase retrieval for real-time phase imaging, parallel computation using sophisticated Graphics Processing Units (GPUs) or Field Programmable Gate Arrays (FPGAs) has been implemented to accelerate phase unwrapping~\cite{chen2017efficient, pham2011off}. However, specialized programming is required that hinders its generalization. Furthermore, obtaining the calibration interferogram for aberration compensation becomes particularly challenging when imaging dense samples, as a sample-free region may not be readily available.

In recent years, artificial neural networks (ANNs), including the widely used, U-Net have achieved simultaneous phase retrieval and elimination of the need for aberration compensation~\cite{wang2024use, Chang2020Cali, yao2020deep, ynet}. Despite simplifying the imaging operation in off-axis QPI, further applying these ANNs for real-time phase imaging is potentially limited by the relatively large computational latency. It is known that the network inference accuracy and latency heavily depend on the network's architecture. Therefore, we need a strategy to identify an optimal network architecture that minimizes computational latency and maintains high accuracy for phase retrieval.

Neural architecture search (NAS)~\cite{NASReview} is a technique to automatically find an optimal network architecture from a large architecture search space. NAS-generated networks have outperformed manually designed networks in many tasks, including classification~\cite{ren2021comprehensive}, semantic segmentation~\cite{autodeeplab}, etc. SparseMask~\cite{sparsemask} is an end-to-end semantic segmentation NAS scheme. SparseMask has a network architecture search space that covers different skip connection strategies from the encoder to the decoder, which enables searching for optimal ways to fuse low-level features rich in spatial details and high-level features containing semantic information. In SparseMask, a differentiable NAS search strategy is used to relax the architecture search space from discrete to continuous, and the gradient descent is used to efficiently search for optimal skip connections. Taking both segmentation accuracy and connectivity sparsity into account, SparseMask attains a similar mean Intersection-over-Union (mIoU) but executes over three times faster, when compared with the widely-used Pyramid Scene Parsing Network (PSPNet) on the semantic segmentation PASCAL VOC 2012 test dataset.

\begin{figure}[tbp]
\centering
\includegraphics[width=0.9\linewidth]{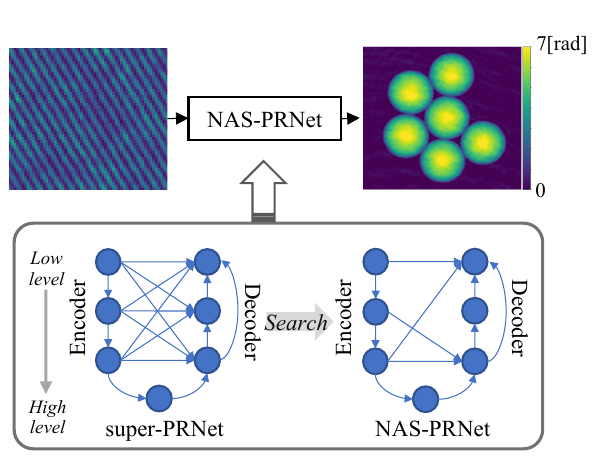}
\vspace{-0.5cm}
\caption{Illustration of searching NAS-PRNet from a neural architecture search space through constructing the super-PRNet. Blue circles represent the encoder and decoder stages, lines with arrows represent the connections from head stages to tail stages. }
\vspace{-0.5cm}
\label{fig:1}
\end{figure}

To attain high-fidelity real-time phase imaging in off-axis QPI, we propose NAS-generated Phase Retrieval Net (NAS-PRNet) as illustrated in Fig.~\ref{fig:1}. The architecture of NAS-PRNet is derived through a customized adaptation of the SparseMask NAS algorithm. We then evaluate NAS-PRNet’s performance in phase retrieval and compare it with the classic U-Net~\cite{unet}\revision{, the most recent Mamba-UNet}~\cite{wang2024mamba}\revision{ and EMCAD}~\cite{rahman2024emcad}\revision{, and original SparseMask to demonstrate its high accuracy and efficiency.} Source code will be available at https://github.com/shuxin626/NAS-Phase-Retrieval-Net.

\vspace{-0.4cm}
\section{Methodology}

To obtain the architecture of NAS-PRNet, an intermediate super-network for phase retrieval (super-PRNet), containing all possible encoder and decoder connections, is first constructed. Each connection in super-PRNet has an uniformly initialized weight of $0.5$. Then, the super-PRNet is trained, and its encoder and decoder connections are pruned, according to the trained connection weights. After that, the architecture of NAS-PRNet is obtained. Finally, the NAS-PRNet is trained, and its performance is evaluated for phase retrieval. Both super-PRNet and NAS-PRNet are trained and tested with the same interferogram dataset. Here, we use biological cells to construct a dataset with diverse features and sizes for generalization. To enlarge the search space to include more possible network architectures, we modified the original SparseMask by loosening its constraints in two aspects: (1) allowing encoder features to propagate into all stages of the decoder, instead of only the corresponding lower-level decoder stages; and (2) applying a global sparsity loss to minimize the total number of connections, instead of a sparsity restriction for each decoder stage on a fixed quantity of connections. Furthermore, to fit SparseMask to the phase retrieval task, we modify its output layer (i.e., adopting a regression head), feature fusing style, kernel size in convolution, and feature depth strategy.
\vspace{-0.3cm}
\subsection{\revision{Structure} of super-PRNet}

\begin{figure}[t]
\centering
\includegraphics[width=0.95\linewidth]{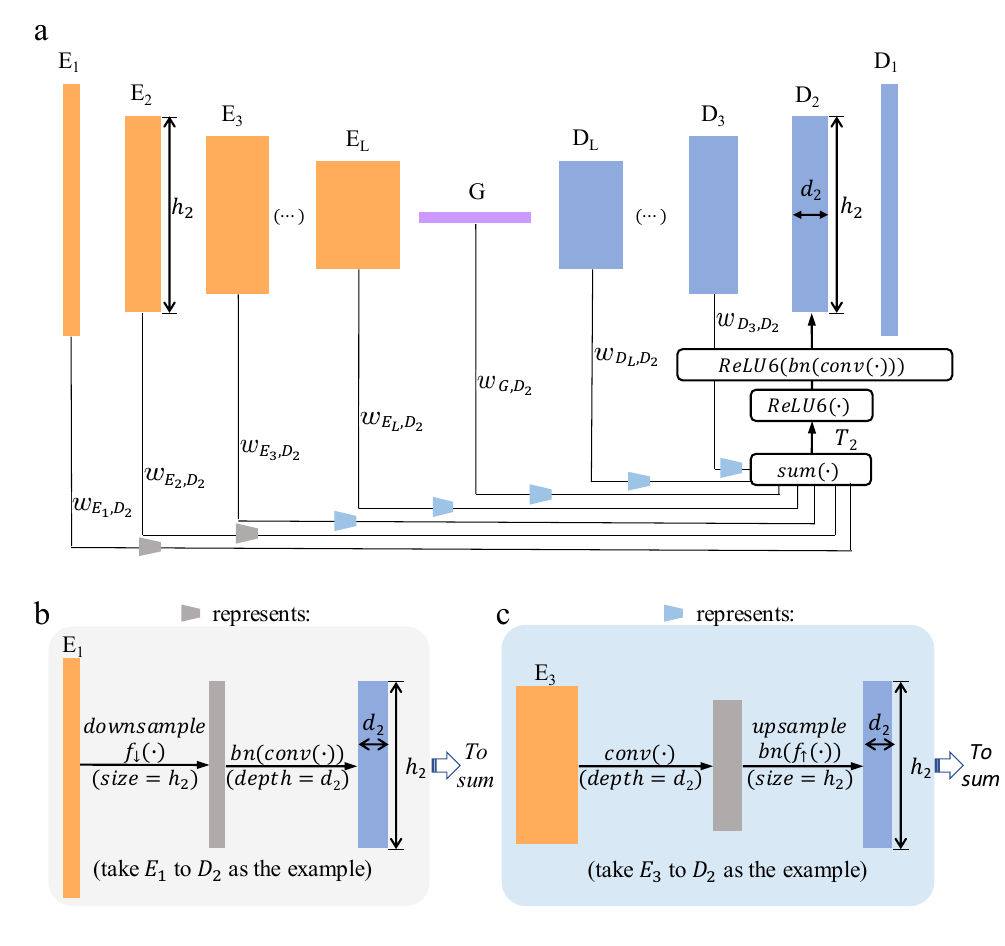}
\caption{Feature concatenation in super-PRNet. (a) Formation of $D_2$. $E_i$: $i_{th}$ encoder feature; $D_j$: $j_{th}$ decoder feature. (b) Processing of input features with spatial sizes larger than $h_2$. (c) Processing of input features with spatial sizes smaller than $h_2$.}
\vspace{-0.5cm}
\label{fig:2}
\end{figure}

The structure of super-PRNet is illustrated in Fig.~\ref{fig:2}, where MobileNet-v2 is implemented as the encoder to efficiently extract multi-level features from an input interferogram. In Fig. 2(a), the encoder features at multiple levels are denoted as $E_l$, where $l$ is the stage index ranging from $1$ to $L=8$ (i.e., 8 encoder features are input to the decoder). In addition, a ground encoder feature G is obtained by applying an average pool with a target size of $3\times3$, and then input to the decoder. The decoder integrates all possible connections between encoder and decoder stages, as well as different stages inside the decoder. Being symmetric with the encoder, the number of stages in the decoder is also $L$. $D_l$ denotes the $l_{th}$ decoder stage feature that has the same feature spatial size of $h_l$ as $E_l$. The feature depth of $D_l$ is set as $min(256,8l)$. For the $l_{th}$ encoder stage, its input features are $\{E_i|1\leq i \leq L\}$, $G$, and $\{D_j|l<j\leq L\}$. As these input features differ in size and depth, we will efficiently fuse them in the following way: (i) for features $m\in M_l^+=\{E_i|1\leq i<l\}$ with spatial size larger than $h_l$, a bilinear down-sampling with a target spatial size of $h_l$ is first applied before a convolution operation with a target feature depth of $d_l$; (ii) for features $m\in M_l^-=\{E_i|l \leq i \leq L\} \bigcup G \bigcup \{D_j|l<j \leq L\}$ with spatial size smaller than $h_l$, a convolution with a target feature depth of $d_l$ is first applied before a bilinear up-sampling with a target spatial size of $h_l$; (3) we fuse the processed input features of the same depth and size as a fused feature $T_l$ by a weighted sum:   
\begin{equation}
\begin{split}
    T_l \!=\!\!\!\!\sum_{m \in M_l^+}\!\!\!\!{w_{m,l} bn(conv(f_{\downarrow}(m)))} \!+\!\!\!\! \sum_{m \in M_l^-}\!\!\!\!{w_{m,l} bn(f_{\uparrow}(conv(m)))}
\end{split}
\label{eq:1}
\end{equation}
where $f_{\downarrow}()$ and $f_{\uparrow}()$ denote the bilinear down-sampling and bilinear up-sampling, respectively; $conv()$ is the 2D convolution with $3\times3$ kernel size; $w_{m,l}$ is the weight of the connection from input feature $m$ to decoder stage feature $D_l$; and $bn()$ represents the batch normalization. $w_{m,l}=0$ indicates the connection does not exist, while $w_{m,l}=1$ indicates the connection exists. $bn()$ ensures the output value of $conv(f_{\downarrow}())$ or $f_{\uparrow}(conv())$ not affecting the connection importance represented by the summation weight $w_{m,l}$. Finally, decoder stage feature $D_l$ is obtained as $D_l = ReLU6(bn(conv(ReLU6(T_l))))$ where $ReLU6()$ is the nonlinear activation function. 

To output a phase map with continuous phase values and the same image size as the input interferogram in a single channel, a regression head is added to $D_1$, i.e., the end feature of the decoder. The regression head is comprised of three consecutive functions: (i) 2D convolution to compress the feature depth of $D_1$ from $8$ to $1$; (ii) interpolation to make the feature spatial size of $D_1$ identical to the input interferogram (originally $h_1$ is only half the size of the input interferogram); (iii) $ReLU6$ nonlinear activation and dividing the obtained value by $6$ to make the output have pixel value ranging from $0$ to $1$ for training.

\vspace{-0.3cm}
\subsection{Deriving NAS-PRNet from Super-PRNet}
The search for the optimal encoder and decoder connections for NAS-PRNet is formulated as a problem of finding the optimal binary subset of the weight set $W =\{w_{m,l}|m\in M_l^+ \bigcup M_l^-, 1 \leq l \leq L\}$. Considering the tradeoff between efficiency and accuracy, the optimization objectives include: (i) making the connectivity as sparse as possible to reduce the computation latency of this network; and (ii) decreasing the phase retrieval loss as much as possible. As it is computationally inefficient to search $W$ in a discrete search space, we relax all weights $w\in W$ to be continuous, ranging from $0$ to $1$, to allow for gradient descent to optimize the connection weights and conduct the architecture search. The synthesized loss function $Loss_*$ used in the training process is formulated as:
\begin{equation}
     Loss_{*} = Loss_{t} + \alpha Loss_b + \beta Loss_s,
\end{equation}

where $Loss_t$ is the phase reconstruction loss which is calculated as the Mixed Gradient Error (MixGE) between the ground truth and the network output as defined in~\cite{van2021optimized}; $Loss_b$ and $Loss_s$ are the binary loss and the network sparsity loss, respectively; and $\alpha$ and $\beta$ are the coefficients for $Loss_b$ and $Loss_s$. $Loss_b$ and $Loss_s$ are defined as: 
\begin{subequations}
\begin{align}
Loss_b = &\frac{\sum_{w^* \in W}\!(-w^*log(w^*)\!-\!(1\!-\!w^*)log(1\!-\!w^*))}{len(W)},\\ 
Loss_s = &\sum_{w^* \in T}w^*
\end{align}
\label{eq:3}
\end{subequations}

In Eq.~\ref{eq:3}a, the loss term $-wlog(w)-(1-w)log(1-w)$ in binary loss will push the weight \revision{$w$} close to $0$ or $1$ during the network training process~\cite{sparsemask}, and $len()$ takes the length of $W$. The mean of all weights $w\in W$ serves as the sparse loss as described in Eq.~\ref{eq:3}b. A smaller $Loss_s$ indicates more weight $w$ values are closer to $0$, namely the connectivity in super-PRNet will be sparser. After tuning, we set $\alpha=5\times10^{-3}$ and $\beta=5\times10^{-5}$.   

\begin{figure}[tbp]
\centering
\includegraphics[width=0.99\linewidth]{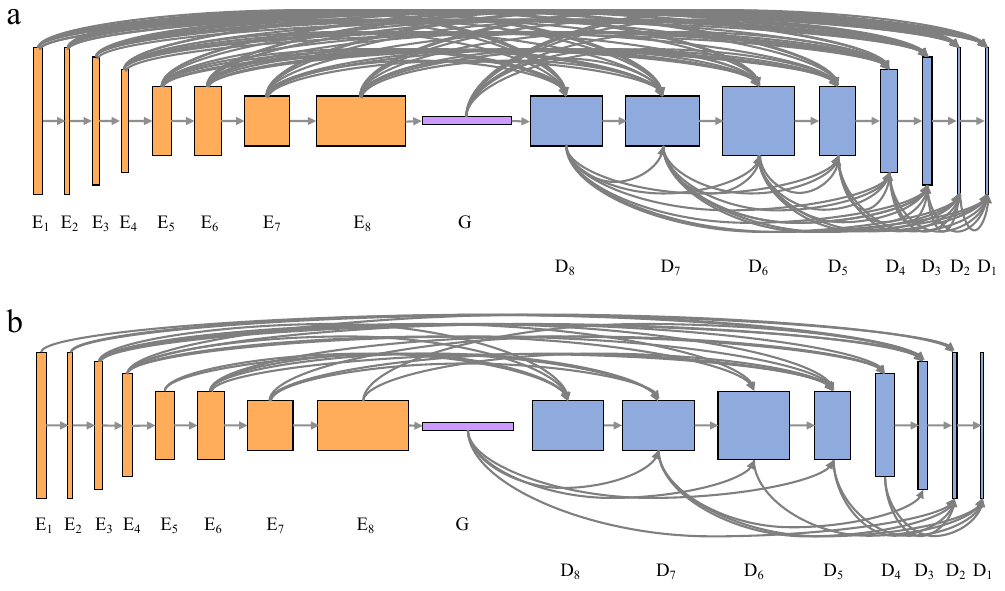}
\vspace{-0.5cm}
\caption{(a) Connections in super-PRNet. (b) Connections in NAS-PRNet. Orange, blue, and purple rectangles represent encoder stages (E), decoder stages (D), and the ground stage (G), respectively.}
\vspace{-0.5cm}
\label{fig:3}
\end{figure}

We first trained the super-PRNet on an NIH/3T3 dataset (for dataset and training details, refer to Supplementary Material). \revision{The training process took $4$ hours on a Supermicro GPU server (Intel Xeon Silver 4210R CPU [}$\times2$\revision{], Nvidia RTX A6000 48GB [}$\times1$\revision{]).} Then, to prune super-PRNet to get NAS-PRNet (Fig.~\ref{fig:3}), we selected the connection weight set \revision{$W$} in the checkpoint with the best validation PSNR and SSIM. The pruning rules are as follows: (i) drop all the connections with weights $w < 0.001$, and (ii) drop all decoder stages without any input features, as well as whose features that are not used by any decoder stages. After pruning, the number of connections in super-PRNet (Fig.~\ref{fig:3}(a)) has been significantly reduced from $100$ to $42$, i.e. reducing the connections by $58\%$, to achieve NAS-PRNet that has a sparse connectivity (Fig.~\ref{fig:3}(b)). In NAS-PRNet, as both low-level and high-level features take part in the formation of each decoder, an optimal fusion of these features can ensure accurate phase retrieval.

In addition to applying the found connection scheme, NAS-PRNet is modified from super-PRNet in producing the fused features by removing $w$ and $bn()$ in Eq.~\ref{eq:1}. Therefore, the fused features $T_l'$ of NAS-PRNet are:
\begin{equation}
    T_l \!=\! \sum_{m \in M_l^+}{w_{m,l} conv(f_{\downarrow}(m))} \!+\! \sum_{m \in M_l^-}{w_{m,l} f_{\uparrow}(conv(m))}             
\end{equation}
\vspace{-0.6cm}
\subsection{Evaluation of NAS-PRNet}

To evaluate the phase retrieval accuracy, we trained NAS-PRNet following the same protocol as super-PRNet but only used the phase reconstruction loss $Loss_t$. We tested the phase retrieval time on a \revision{mid-range Lenovo} laptop (Intel i7-9750H CPU [$\times1$], Nvidia GeForce RTX 2060 6GB [$\times1$]) using full NIH/3T3 images with size of $1024 \times 1024$ pixels. 

\vspace{-0.2cm}
\section{Results}
\begin{figure}[tbp]
\centering
\includegraphics[width=\linewidth]{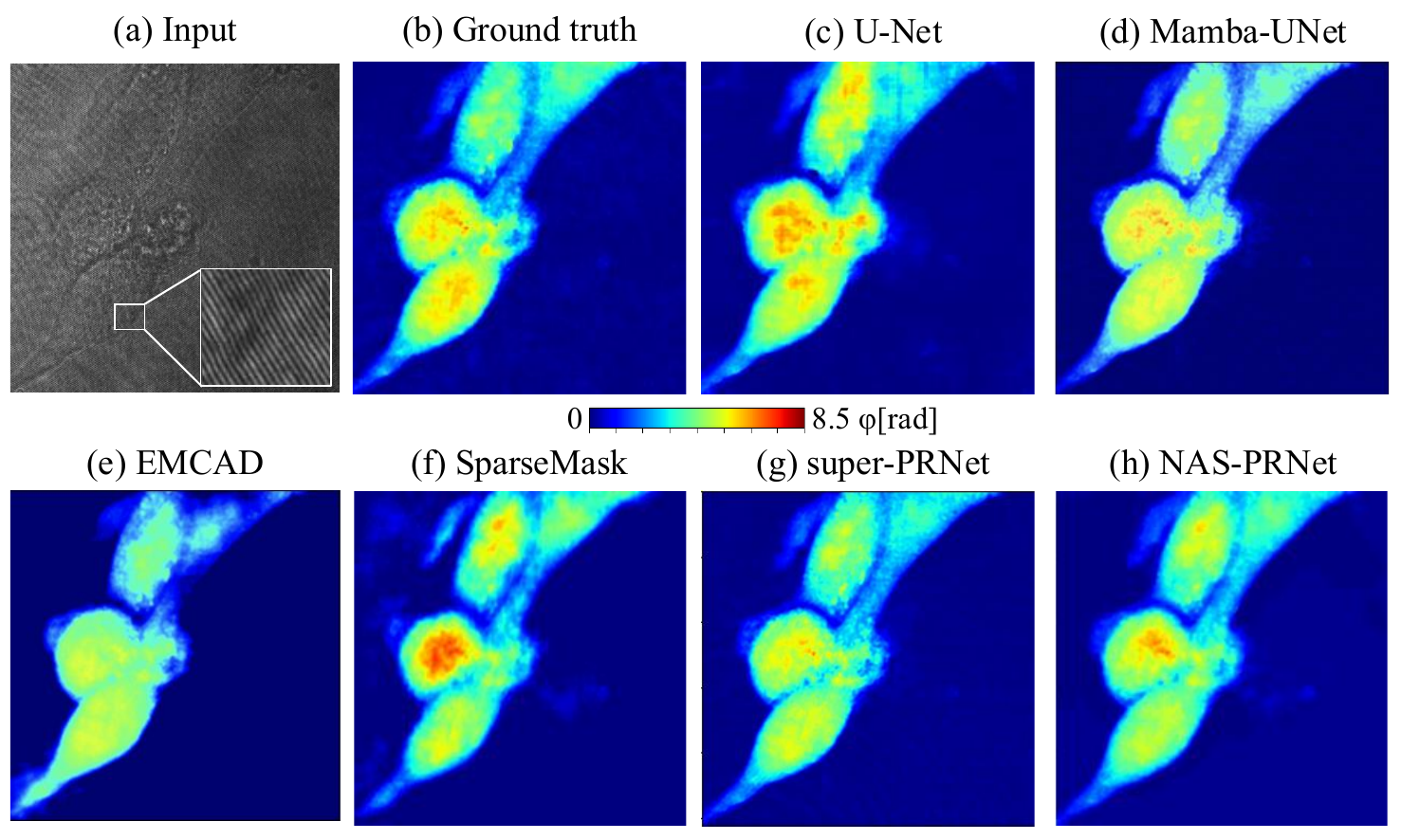}
\caption{Comparison of super-PRNet and NAS-PRNet with baseline methods on the NIH/3T3 cell dataset. (a) Input interferogram. (b) Ground truth. (c) U-Net. (d)Mamba-UNet. (e) EMCAD. (f) SparseMask. (g) super-PRNet. (h) NAS-PRNet.}
\vspace{-0.5cm}
\label{fig:4}
\end{figure}

\revision{We compare the performance of super-PRNet and NAS-PRNet against several baseline methods, including the classic U-Net}~\cite{unet}\revision{, the recent Mamba-UNet and EMCAD based on popular Mamba and attention mechanisms from natural language processing, respectively}~\cite{wang2024mamba, rahman2024emcad}\revision{, as well as SparseMask (implementation details can be found in Supplementary Material). We present the testing results using the NIH/3T3 dataset on a mid-range laptop in Tab.}~\ref{tab:1}\revision{ and Fig.}~\ref{fig:4}\revision{. To analyze the results, we divide the baseline methods into heavy-weight (U-Net and Mamba-UNet) and light-weight (EMCAD and SparseMask) groups. The heavy-weight group achieves higher accuracies ($36.8-37.8$ dB PSNR) but significantly slower inference speed ($373-481$ ms), while the light-weight group offers faster inference speed ($21-88$ ms) with lower accuracy ($30.1-31.8$ dB PSNR). Note that the longer inference time of Mamba-UNet with lower FLOPs is likely due to less optimized GPU implementations compared to U-Net’s well-established convolution operators. In contrast, NAS-PRNet achieves an optimal balance between accuracy ($36.7$ dB PSNR) and efficiency ($31$ ms latency), thus demonstrating comparable performance with $12\times$ speedup over U-Net and slightly lower performance with $15\times$ speedup over Mamba-UNet.} Moreover, the comparable accuracy between NAS-PRNet and super-PRNet demonstrates that our searching strategy is effective, as it has selectively pruned redundant connections and preserved critical connections to maintain a high phase retrieval accuracy. To test the robustness of NAS-PRNet, we used a white blood cell dataset~\cite{shu2021artificial} acquired from a different off-axis QPI system and achieved a high phase retrieval accuracy (PSNR $44.0$ dB and SSIM $93.4\%$) that is comparable to U-Net (refer to Supplementary Material for details).

Note that when using the traditional Fourier transform-based phase retrieval method (i.e., the ground truth map), phase unwrapping is required after obtaining the calibrated phase map. Using the Goldstein algorithm, phase unwrapping takes $528$ ms for an image size of $1024 \times1024$ \revision{when executed on the CPU of the laptop}. In contrast, our NAS-PRNet outputs the unwrapped phase in just $31$ ms \revision{on the same laptop using GPU}, which is over $17$ times faster. The phase unwrapping range of NAS-PRNet is currently limited to $0$ – $12$ rad (the phase range in the dataset). However, this range can be extended by training the network with datasets containing larger phase values. Moreover, NAS-PRNet can automatically correct system aberration without using a calibration phase map, which significantly simplifies the phase imaging experiments.

\begin{table}[t]
    \centering
    \caption{\label{PhaseRetrievalPerformanceComparison} Phase retrieval performance comparison}
    \begin{tabular*}{\linewidth}{@{}l*{5}{@{\extracolsep{0pt plus 12pt}}l}}
        \hline
        Methods     & PSNR   & SSIM   & Params & \revision{FLOPs} &\makecell{Latency} \\
        \hline
        U-Net        & $36.8$ dB & $87.4\%$ & $37.7$M & \revision{$971.4$G} & $373$ ms   \\
        \revision{Mamba-UNet} & \revision{$37.8$ dB} & \revision{$87.8\%$} & \revision{$19.1$M} & \revision{$94.6$G} & \revision{$481$ ms}\\
        \revision{EMCAD} & \revision{$30.1$ dB} & \revision{$60.4\%$}& \revision{$3.9$M}& \revision{$30.2$G} & \revision{$88$ ms}\\
        SparseMask  & $31.8$ dB & $65.6\%$ & $1.8$M & \revision{$6.9$G} & $21$ ms    \\
        \hline
        \revision{super-PRNet (ours)}  & \revision{$37.6$ dB} & \revision{$85.6\%$} & \revision{$7.2$M} & \revision{17.5G} & \revision{$60$ ms}
        \\
        NAS-PRNet (ours)        & $36.7$ dB & $86.6\%$ & $4.4$M & \revision{$11.3$G} & $31$ ms   \\
        \hline
    \end{tabular*}
    \label{tab:1}
    \vspace{-0.4cm}
\end{table}

\vspace{-0.4cm}
\section{Conclusion}
In conclusion, we have developed NAS-PRNet for phase retrieval and optimized its architecture to balance the output accuracy and inference speed. \revision{Compared with the recent Mamba-UNet, NAS-PRNet reduces inference time by $15\times$ with only slightly lower phase retrieval accuracy.} With the high phase retrieval speed offered by NAS-PRNet and the single-shot capability of off-axis QPI, one may demonstrate many real-time imaging applications, such as profiling the morphology of living cells and quantifying their dynamics. The current search space of NAS-PRNet is only limited to the connection scheme, but it could be further expanded to cover layer depth, layer manipulation, and so on, which may lead to the discovery of a more efficient network architecture. 
\vspace{-0.5cm}
\bibliographystyle{IEEEtran}
\bibliography{nas}
\end{document}